\documentclass[conference]{IEEEtran}
\usepackage{graphicx}
\usepackage{cite}
\usepackage{picinpar}
\usepackage{amsmath}
\usepackage{amsfonts}
\usepackage{bm}
\usepackage{url}
\usepackage{flushend}
\usepackage[latin1]{inputenc}
\usepackage{colortbl}
\usepackage{soul}
\usepackage{multirow}
\usepackage{booktabs}
\usepackage{threeparttable}
\usepackage{subcaption}
\usepackage{pifont}
\usepackage{alltt}
\usepackage{enumerate}
\usepackage{siunitx}
\usepackage{breakurl}
\usepackage{epstopdf}
\usepackage{pbox}

\usepackage{algorithmicx}
\usepackage{algorithm}
\usepackage{algpseudocode}  
\algnewcommand\algorithmicinput{\textbf{Input:}}
\algnewcommand\Input{\item[\algorithmicinput]}

\algnewcommand\algorithmicoutput{\textbf{Output:}}
\algnewcommand\Output{\item[\algorithmicoutput]}
\algnewcommand\True{\textbf{true}}
\algnewcommand\False{\textbf{false}}

\algnewcommand{\LineComment}[1]{\Statex \(\triangledown \) #1}  

\algnewcommand{\IfThenElse}[3]{
	\State \algorithmicif\ #1\ \algorithmicthen\ #2\ \algorithmicelse\ #3 \algorithmicend}
\algnewcommand{\IfThen}[2]{\State \algorithmicif\ #1\ \algorithmicthen\ #2 \algorithmicend}

\usepackage[colorlinks]{hyperref}
\usepackage{xcolor}
\hypersetup{
	linkcolor={red!60!black},
	citecolor={blue!50!black},
	urlcolor={blue!80!black}
}

\graphicspath{{./}}

\newcommand{\bth}{\bm{\theta}}
\newcommand{\bx}{\bm x}
\newcommand{\bbR}{\mathbb R}
\definecolor{rvblue}{rgb}{0.03, 0.27, 0.49} 

\newcommand{\RNum}[1]{\uppercase\expandafter{\romannumeral #1\relax}}

\begin{document}
	\title{On Solar Photovoltaic Parameter Estimation: Global Optimality Analysis and a Simple Efficient Differential Evolution Method}
	
	


			

\makeatletter
\newcommand{\linebreakand}{%
 \end{@IEEEauthorhalign}
 \hfill\mbox{}\par
 \mbox{}\hfill\begin{@IEEEauthorhalign}
}
\makeatother

	\author{\IEEEauthorblockN{Shuhua Gao}
	\IEEEauthorblockA{\textit{School of Control Science and Engineering} \\
	\textit{Shandong University}\\
	Jinan, Shandong, China\\
	shuhuagao@sdu.edu.cn}
	\and
	\IEEEauthorblockN{Yunyi Zhao}
	\IEEEauthorblockA{\textit{Electrical \& Computer Engineering} \\
	\textit{National University of Singapore}\\
	Singapore \\
	e0679990@u.nus.edu}
	\and
	\IEEEauthorblockN{Cheng Xiang}
	\IEEEauthorblockA{\textit{Electrical \& Computer Engineering} \\
	\textit{National University of Singapore}\\
	Singapore \\
	elexc@nus.edu.sg}
	\and
	\IEEEauthorblockN{Ming Yu}
	\IEEEauthorblockA{\textit{Power Automation Pte Ltd} \\
	Singapore \\
	ming.yu@pa.com.sg}
	\and
	\IEEEauthorblockN{Kuan Tak Tan}
	\IEEEauthorblockA{\textit{Engineering Cluster} \\
	\textit{Singapore Institute of Technology}\\
	Singapore \\
	KuanTak.Tan@SingaporeTech.edu.sg}
	\and
	\IEEEauthorblockN{Tong Heng Lee}
	\IEEEauthorblockA{\textit{Electrical \& Computer Engineering} \\
	\textit{National University of Singapore}\\
	Singapore \\
	eleleeth@nus.edu.sg}
	}

\maketitle

\begin{abstract}
	A large variety of sophisticated metaheuristic methods have been proposed for photovoltaic parameter extraction. Our aim is not to develop another metaheuristic method but to investigate two practically important yet rarely studied issues: (i) whether existing results are already globally optimal; (ii) whether a significantly simpler metaheuristic can achieve equally good performance. We take the two widely used I-V curve datasets for case studies. The first issue is addressed using a branch and bound algorithm, which certifies the global minimum rigorously or locates a fairly tight upper bound, despite its intolerable slowness. These values are useful references for fair evaluation and further development of metaheuristics. Next, extensive examination and comparison reveal that, perhaps surprisingly, an elementary differential evolution (DE) algorithm can either attain the global minimum certified above or obtain the best-known result. More attractively, the simple DE algorithm takes only a fraction of the runtime of state-of-the-art metaheuristic methods and is particularly preferable in time-sensitive applications. This novel, unusual, and notable finding also indicates that the employment of increasingly complicated metaheuristics might be somewhat overkilling for regular PV parameter estimation. Finally, we discuss the implications of these results for future research and suggest the simple DE method as the first choice for industrial applications.
\end{abstract}


\markboth{}%
{}

\definecolor{limegreen}{rgb}{0.2, 0.8, 0.2}
\definecolor{forestgreen}{rgb}{0.13, 0.55, 0.13}
\definecolor{greenhtml}{rgb}{0.0, 0.5, 0.0}

\section{Introduction}\label{sec: intro}
Accurate modeling of solar photovoltaic (PV) systems is necessary for their effective design, simulation, power forecasting, and optimal control \cite{jordehi2016ParameterEstimation,   li2019ParameterExtraction, yang2020ComprehensiveOverview}. The dominating method to describe solar PV systems uses an analogous electrical circuit model \cite{villalva2009ComprehensiveApproach}, which has been further specialized to the single-diode model (SDM), the double-diode model (DDM). Despite the intuitiveness of these circuit models,  the main difficulty lies in the accurate determination of unknown parameters in the model \cite{chenouard2020IntervalBranch, yang2020ComprehensiveOverview, chin2015CellModelling, villalva2009ComprehensiveApproach}. 

PV parameter estimation is commonly formulated as a nonlinear optimization problem from the perspective of I-V (current-voltage) curve fitting. The problem has been widely attempted with various metaheuristic algorithms.  Most metaheuristics are population-based by exploiting a swarm of interacting agents to search the solution space efficiently \cite{yang2020ComprehensiveOverview}. Since metaheuristic algorithms are mostly not problem-specific, any metaheuristic optimizer may be applied to PV parameter estimation in principle. It is unsurprising that a large number of metaheuristic methods have been proposed for PV parameter estimation. Some recent examples include guaranteed convergence particle swarm optimization \cite{nunes2018NewHigha},  improved JAYA optimization \cite{yu2017ParametersIdentification},  performance-guided JAYA \cite{yu2019PerformanceguidedJAYA}, self-adaptive ensemble-based differential evolution \cite{liang2020ParametersEstimation}, teaching-learning-based optimization \cite{li2019ParameterExtraction, chen2018TeachingLearning},  grey wolf optimizer and cuckoo search based hybrid method \cite{long2020NewHybrid}, among many others.  We omit their technical details due to space limit. Interested readers may refer to \cite{chin2015CellModelling, yang2020ComprehensiveOverview, jordehi2016ParameterEstimation} for detailed reviews.

Despite the increasing interest in such metaheuristics, none of them can guarantee or identify the discovery of the global optimum \cite{chenouard2020IntervalBranch}. Moreover, the minimal root mean square error (RMSE) of curve fitting attained by different metaheuristics has suffered from stagnation with no further reduction in recent studies (see  \cite[Table 3]{li2019ParameterExtraction},   \cite[Table 3]{yu2019PerformanceguidedJAYA}, and Table \ref{tbl: compare}). Thus, one may wonder naturally whether the best-known result is already the global minimum such that we can avoid futile efforts by building more advanced metaheuristics blindly.  Also, since a variety of metaheuristics can get the same RMSE value, another natural query is how sophisticated a metaheuristic has to be to achieve effective PV parameter estimation. An industrial practitioner desires certainly an effective yet simple and fast algorithm. In particular, the algorithm's efficiency is critical for time-sensitive applications, for example, the real-time monitoring of solar cell degradation via photovoltaic curves telemetry using a microprocessor on a satellite \cite{gutierrez2018SystemonChipRealTime}.

In this study, we attempt to answer the above two questions through extensive investigations using the two most broadly studied benchmark datasets (mainly to facilitate comparison) \cite{easwarakhanthan1986NonlinearMinimizationa}. 
The main contributions of this paper are listed below.
\begin{itemize}
	\item The global minimum RMSE of the SDM has been certified on both datasets for the first time using an interval arithmetic based branch and bound method.
	Besides, a useful upper bound of the global minimum for the DDM is obtained. These values can serve as valuable references for the assessment and  development of metaheuristics. 
	\item We show that an \textit{intentionally} simple differential evolution (DE) algorithm is adequate to attain the global minimum for the SDM and achieve equally high accuracy for the DDM compared with a variety of sophisticated metaheuristics. Moreover, the DE algorithm stands out with high performance stability and incomparable time efficiency thanks to its simplicity, which renders itself particularly suitable for real-time applications.
	\item Based on our findings and comparison with state-of-the-art methods, we recommend the simple DE  to solar industry engineers as the first choice in practical applications, especially time-sensitive ones. Besides, we provide useful suggestions for researchers to refresh viewpoints on PV parameter estimation and to refrain from possible over-engineering in designing overcomplicated metaheuristics. 
\end{itemize}

Finally, we would like to emphasize that, as implied in the above contributions, the purpose of this paper is definitely \textit{not} to develop yet another new metaheuristic method for PV parameter estimation. 
The main objective of  this study is to reveal rigorously the limit of estimation accuracy that a metaheuristic method can achieve and to revive the simple classic differential evolution algorithm for PV parameter estimation that has been overlooked hastily in the current literature.

The remainder of this paper is organized as follows. Common PV models are first introduced in Section \ref{sec: model}, followed by the optimization problem formulation. The two optimization methods are described in Section \ref{sec: methods}. We then apply the two methods to two benchmark datasets, report the results, and conduct a detailed comparison in Section \ref{sec: results}. Finally, we conclude this paper with Section \ref{sec: conclusion}. 

\section{System modeling and problem formulation}\label{sec: model}
\subsection{Modeling of PV systems}
The electrical circuit corresponding to the SDM is shown in Fig.\ \ref{fig: SDM-DDM}(a). Specifically, the circuit contains a current source $ I_{ph} $, which refers to the photocurrent generated by the PV cell, a diode flowing current $ I_d $, and two resistors with resistance $ R_{p} $ and $ R_s $, respectively. We can calculate the diode current $ I_d $ using the Shockley equation as follows,
\begin{equation} \label{eq: SM-Id}
	I_d = I_0 \left[\exp\left(\frac{q(V + IR_s)}{nkT} \right)  - 1  \right],
\end{equation}
where $ I_0 $ is the reverse saturation current of the diode, $ n $ is the diode ideal factor, $ T $ is the temperature in Kelvin, and $ V $ is the output voltage of the cell. The other terms are just physical constants: the electron charge $ q =  1.60217646\times 10^{-19}$ C and the Boltzmann constant  $ k =  1.3806503  \times 10^{-23}$ J/K. 

The output current $ I $ is computed using first principles by 
\begin{equation} \label{eq: SM}
	I = I_{ph} -  I_0 \left[\exp\left(\frac{q(V + IR_s)}{nkT} \right)  - 1  \right] - \frac{V + IR_s}{R_p}.
\end{equation}

There are five unknown parameters in \eqref{eq: SM}, which are collected into a parameter vector $ \bm{\theta}_S = [I_{ph}, I_0, n, R_s, R_p] $. 

Despite the simplicity and usefulness of the above SDM, it does not consider the effect of recombination current loss in the depletion region \cite{jordehi2016ParameterEstimation, chin2015CellModelling}. An additional diode can be introduced into the circuit to compensate for this specific loss to attain higher accuracy. The equivalent circuit of the DDM is illustrated in Fig.\ \ref{fig: SDM-DDM}(b).
In analogy to the SDM \eqref{eq: SM}, the DDM is derived straightforwardly as follows:
\begin{align} \label{eq: DDM}
	&I &= &I_{ph} - I_{d1} - I_{d2} - I_{p} \nonumber\\
	&&= &I_{ph} -  I_{01} \left[\exp\left(\frac{q(V + IR_s)}{n_1kT} \right)  - 1  \right]  -  \nonumber\\
	&& & I_{02} \left[\exp\left(\frac{q(V + IR_s)}{n_2kT} \right)  - 1  \right] - \frac{V + IR_s}{R_p},
\end{align}
where $ I_{01} $ and $ I_{02} $ are the reverse saturation current of the two diodes, and $ n_1 $ and $ n_2 $ denote the ideality factor of the two diodes, respectively.
The DDM has seven parameters in total, denoted by $ \bm{\theta}_D = [I_{ph}, I_{01},  I_{02}, n_1, n_2, R_s, R_p] $. 

\begin{figure}[tb]
	\centering
	\includegraphics[width=80mm]{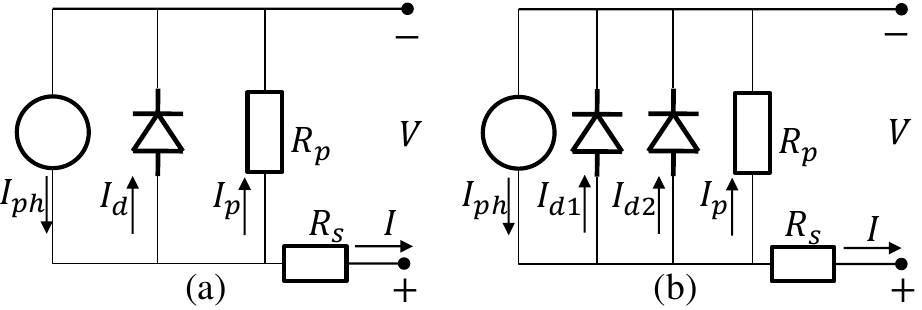} 
	\caption{Equivalent circuit of a PV cell. (a) SDM; (b) DDM.}
	\label{fig: SDM-DDM}
	\vspace{-2mm}
\end{figure} 

A PV module contains multiple PV cells connected in series or parallel. It is standard to assume the same parameter values for all cells for computational tractability purposes. Then, all cells are lumped into a single, functionally equivalent cell \cite{easwarakhanthan1986NonlinearMinimizationa, yu2019PerformanceguidedJAYA,liang2020EvolutionaryMultitask}. Thus, the same procedure is used to fit either the SDM \eqref{eq: SM} or the DDM \eqref{eq: DDM} to the I-V curve of a PV module.

\subsection{Optimization problem formulation} \label{sec: problem}
The fundamental principle of parameter estimation via I-V curve fitting is to find appropriate parameter values such that the current values calculated with either the SDM \eqref{eq: SM} or the DDM \eqref{eq: DDM} match  the measurement values for a set of data points \cite{jordehi2016ParameterEstimation}. 
Without loss of generality, we discuss below the problem formulation using the SDM \eqref{eq: SM}, termed $ f_S $ hereafter, with parameters $ \bth_S $, whose principle can be transplanted to the DDM case seamlessly.

Note that we cannot write down a simple closed-form solution $ I = f_S^{-1}(V) $ for the model $ f_S $ \eqref{eq: SM} to compute $ I $ given $ V $. As a workaround, given a measurement $ (V^m, I^m) $ and a tentative parameter vector $ \bm{\theta}_S $, the majority of metaheuristic-based studies compute the \textit{predicted} current with the model approximately but computationally economically as
\begin{equation} \label{eq: V->I}
	I \approx f_S(V^m, I^m; \bm{\theta}_S), 
\end{equation}
and try to reduce the deviation between $ I $ and $ I^m $ by adjusting $\bm{\theta}_S  $ (see \cite{li2019ParameterExtraction},  \cite{yu2019PerformanceguidedJAYA}, and \cite{liang2020EvolutionaryMultitask} among others).

The root mean square error (RMSE) is widely used to quantify the difference between computed current values and the measurement values \cite{liang2020ParametersEstimation, gutierrez2018SystemonChipRealTime}. Supposing there are $ N $ data points in the I-V curve,  we get the following constrained optimization problem, which is known widely as \textit{nonlinear least-squares regression} in the literature.
\begin{subequations}\label{eq: problem}
	\begin{align} 
		&\text{minimize}  &\quad &J(\bth) = \sum_{i=1}^{N} \left(f(V^m_i, I^m_i; \bth) - I^m_i \right)^2,  \label{eq: problem obj}\\  
		&\text{subject to} &  &\bth \in \Theta. \label{eq: problem cst}
	\end{align}
\end{subequations}
where $ f $ refers to either the SDM $ f_S $ \eqref{eq: SM} or the DDM $ f_D $ \eqref{eq: DDM}, and $ \bm{\theta} $ is the corresponding parameter vector $ \bm{\theta}_S $ or $ \bm{\theta}_D $.
$ (V^m_i, I^m_i) $ is the $ i $-th data point in measurement. $ \Theta $ denotes the specified bound constraints of $\bm \theta $ that take physical reality into consideration (see Table \ref{tbl: param-range} for examples).

\section{Optimization methods} \label{sec: methods}
In this section, we apply a branch and bound (B\&B) based \textit{deterministic} global optimization technique for rigorous certification of the solution optimality. Then, we choose deliberately a simple \textit{stochastic} optimization algorithm, \textit{differential evolution} (DE), to compete the increasingly sophisticated metaheuristic methods prevalent in the literature.

\subsection{Deterministic global optimization with an interval arithmetic based branch and bound algorithm} \label{sec: BB}
The fundamental task of deterministic global optimization (DGO) is to determine \textit{rigorously} (i.e., with theoretical guarantees) the global minimum of an objective function $ f $ subject to a set of constraints \cite{floudas2013deterministic}. However, finding the global minimum for a general nonconvex optimization problem like \eqref{eq: problem} has been proved to be NP-hard \cite{floudas2013deterministic}. We are practically more interested in identifying a solution sufficiently close to the true global minimum, called  the $ \epsilon $\textit{-global minimum} \cite{floudas2013deterministic}.

The most popular algorithmic framework of DGO is arguably the branch and bound (B\&B) method \cite{floudas2013deterministic}. The search space is divided recursively into smaller subspaces and forms accordingly a tree structure of subproblems. The consequential pruning of search space is performed by eliminating subproblems whose lower bound is no better than the best upper bound found so far.  Interval analysis is a handy tool to estimate the lower and upper bounds of regions/branches of the search space, whose technical details are presented in \cite{hansen2003global}. 
\begin{algorithm}[tb]
	\small
	\caption{Interval branch and bound optimization} \label{alg: BB}
	\begin{algorithmic}[1]
		\Input objective function $ f:  \mathbb{R}^n \rightarrow \bbR$ and bound constraints $ X \subset \bbR^n $, precision parameters $ \epsilon_f $ and $ \epsilon_x $
		\Output lower and upper bounds of the global minimum value $ [\underline{f}, \bar{f}] $, a list of boxes $ L_S $ that contain all possible global minimizers
		\State initialize a list $ L \gets \{[\bx]\} $ with $ [\bx] $ corresponding to $ X $
		\State initialize an empty candidate solution list $ L_S$
		\State $ \bar{f} \gets \infty$  \Comment{Upper bound of $ f^* $}
		\While{$ L \ne \emptyset $}
		\State choose $[\bx]  \in L $ and remove $ [\bx] $ from $ L $  \label{line: box choose}
		\State contract $ [\bx] $ \label{line: box contract}
		\State evaluate $ f $ at the center of $ [\bx] $ and get value $ f_c $
		\State Update $ \bar{f} $ by $ \bar{f} \gets \min\{\bar{f}, f_c \} $
		\If{$ [\bx] $ satisfies criteria \eqref{eq: termination}}  \label{Line: criterion}
		\State append $ [\bx] $ to $ L_S $
		\Else
		\State split $ [\bx] $ into subboxes and add them to $ L $ \label{line: box split}
		\EndIf
		\EndWhile
		\State remove any box $ [\bx] \in L_S $ from $ L_S $ with $ \underline{f}([\bx]) > \bar{f} $  \label{line: pp}
		\State $ \underline{f} \gets \min_{[\bx] \in L_S}{\underline{f}([\bx])}$ \Comment{Lower bound of $ f^* $}
	\end{algorithmic}
\end{algorithm}

The general B\&B framework with interval arithmetic is depicted in Algorithm \ref{alg: BB}, where $ [\bx] $ denotes an $n$-dimensional interval vector (also known as an \textit{interval box}, whose components are intervals). Applying a function $f:  \mathbb{R}^n \rightarrow \bbR$ to $ [\bx] $ yields another interval termed $f([\bx]) = [\underline{f}([\bx]), \bar{f}([\bx])] $, where $\underline{f}([\bx])$ and $ \bar{f}([\bx])$ denote the lower and upper bounds respectively and are calculated rigorously with interval analysis  \cite{hansen2003global}. 
Each iteration is composed of three main components: box selection (Line \ref{line: box choose}), box contracting (Line \ref{line: box contract}), and box splitting (Line \ref{line: box split}).  In particular, the purpose of contracting is to delete subboxes inside $ [\bx] $ that cannot contain a globally optimal solution to reduce search space \cite[Chapter 12]{hansen2003global}. In order to be included in $ L_S $, a box $ [\bx] $ must satisfy two conditions that are checked in Line \ref{Line: criterion}:
\begin{equation} \label{eq: termination}
	\text{width}([\bx]) \le \epsilon_x, \ \text{width}(f([\bx])) \le \epsilon_f,
\end{equation}
where $ \text{width}(\cdot) $ denotes the \textit{width} of a box defined by its largest diameter \cite{floudas2013deterministic}. $\epsilon_x  $ and $ \epsilon_f $ are two tolerance parameters provided by the user, often known as the \textit{precision}. After the main loop finishes, we post-process the solution list $ L_S $ in Line \ref{line: pp} to discard boxes which cannot contain the global minimum $ \bx^* $ according to the latest knowledge of $ \bar{f} $.

At the end of Algorithm \ref{alg: BB}, we get the (usually very tight) bounds of the global minimum $ f^* \in [\underline{f}, \bar{f}] $. It is guaranteed that $\bar{f}([\bx]) - f^* \le 2\epsilon_f , \forall [\bx] \in L_S$ \cite{hansen2003global}. Any $ \bx$ inside the remaining boxes $ L_S $ becomes an $ \epsilon $-global minimum with $ \epsilon=2\epsilon_f $ in this case \cite{floudas2013deterministic}. Though the \textit{exact} global minimum $ \bx^* $ and $ f^* $ are still unknown and remain computationally intractable, a reasonably tight bound by setting small $ \epsilon_f $ and $ \epsilon_x $ in Algorithm \ref{alg: BB} is usually enough for practical purposes. Note that Algorithm \ref{alg: BB} only sketches out the basic skeleton of interval B\&B algorithms.  We resort to a dedicated interval analysis library \texttt{ibexopt} (\url{http://www.ibex-lib.org} (v2.8)) in actual implementation (see \cite{hansen2003global} for details).

\subsection{Stochastic global optimization with a simple DE}\label{sec/simpleDE}
Though an interval B\&B algorithm can ascertain the global optimum rigorously in theory, it is generally much more computationally expensive than metaheuristic algorithms, rendering itself impractical in industrial applications \cite{chenouard2020IntervalBranch}. We will report its intolerably long running time in Section \ref{sec: global}. {In many engineering applications, it is either unnecessary or computationally intractable to obtain the \textit{exact} global minimum \cite{floudas2013deterministic}, and metaheuristic methods are particularly useful in these scenarios.} In view of the abundance of metaheuristics (recall Section \ref{sec: intro}), we take an elementary differential evolution (DE) algorithm \textit{on purpose} to investigate whether highly complicated metaheuristics are really necessary for PV parameter estimation. Again, we would like to emphasize that our objective is not to develop yet another new metaheuristic but to investigate whether a fundamental one is empirically sufficient for practical PV parameter estimation tasks. 

Each iteration of DE comprises three key steps: selection, crossover, and mutation. The distinguishing feature of DE is its mutation with \textit{difference} vectors \cite{das2011DifferentialEvolution}. To minimize a function $ f:  \mathbb{R}^n \rightarrow \bbR$ with bound constraints $ X \subset \bbR^n $, we outline the simple DE in Algorithm \ref{alg: DE}, whose main body includes only five lines of code in agreement with its simplicity. The initial population $ P $ comprises $ N_p $ vectors, and each initial vector $ \bx_i^0, i \in [1, N_p] $ is generated randomly by
\begin{equation} \label{eq: de init}
	x_{i, j}^0 = \underline{b}_j + \text{rand}(0, 1) \cdot (\bar{b}_j - \underline{b}_j),
\end{equation}
where $ x_{i, j}^0 $ denotes the $ j $-th component of $ \bx_i^0 $, $\underline{b}_j$ and $ \bar{b}_j $ represent the lower and upper bound of the $ j $-th variable respectively, $ j\in [1, n] $. Besides, $ \text{rand}(0, 1) $ generates a random number between 0 and 1. 
\begin{algorithm}[tb]
	\small
	\caption{Simple differential evolution} \label{alg: DE}
	\begin{algorithmic}[1]
		\Input objective function $ f:  \mathbb{R}^n \rightarrow \bbR$ and bound constraints $  X \subset \bbR^n $, control parameters $N_p, C_r, F, G $
		\Output the best vector $ \hat{x} $ and the  function value $ \hat{f}$
		\State generate randomly an initial population $ P^0 \gets\{ \bx_i^0 \}_{i=1}^{N_p}$ with \eqref{eq: de init}
		\For{$ g $ from $ 0 $ to $ G -1$} \label{line: for G}
		\For{each vector $ \bx_i^g \in P^g $}
		\State generate a donor vector $ \bm{v}_i^g $ by \eqref{eq: mutation} \Comment{mutation}
		\State $ \bm{v}_i^g \gets \text{bounce-back}(\bm{v}_i^g)$  by \eqref{eq: bback} \label{line: bback}
		\State $ \bm{u}_i^g \gets \text{crossover}(\bm{v}_i^g, \bm{x}_i^g)$ by \eqref{eq: cx}
		\State $ \bx_i^{g+1} \gets \text{select}(\bm{u}_i^g, \bm{x}_i^g)$ by \eqref{eq: selection} \label{line: select}
		\State insert $ \bx_i^{g+1} $ into the new population $ P^{g+1} $
		\EndFor
		\EndFor
		\State $ \hat{x} \gets$ the best vector in $ P^G $ and $ \hat{f} \gets f(\hat{x})$
	\end{algorithmic}
\end{algorithm}

Several mutation strategies have been developed  for DE \cite{das2011DifferentialEvolution}. Here we adopt the most commonly used one called the ``DE/rand/1'' scheme. For each vector in the $ g $-th iteration, a \textit{donor} vector $ \bm{v}_i^g,  i \in [1, N_p],$ is produced  by 
\begin{equation}\label{eq: mutation}
	\bm{v}_i^g = \bx_{a}^g + F(\bx_b^g - \bx_c^g), \  a\ne b \ne c \ne i,
\end{equation}
where three indices $ a, b, c \in [1, N_p] $ are randomly chosen, and $ F $ is  the \textit{scaling factor} typically in the range $ [0.4, 1] $ \cite{das2011DifferentialEvolution}. 

Note that the donor vector $ \bm{v}_i^g $ in \eqref{eq: mutation} may lie outside the bounded region $ X $. We adapt a simple  \textit{bounce-back} strategy \cite{das2011DifferentialEvolution} to handle bound constraints in Line \ref{line: bback}, which relocates each infeasible component between the bound it violates and the corresponding value of the \textit{target} vector $ \bm{x}_i^g $:

\begin{equation} \label{eq: bback}
	v_{i, j}^g \gets \begin{cases}
		\underline{b}_j + \text{rand}(0, 1) \cdot (x_{i, j}^g - \underline{b}_j) \ \text{if } v_{i, j}^g < \underline{b}_j \\
		\bar{b}_j - \text{rand}(0, 1) \cdot (\bar{b}_j  - x_{i, j}^g) \ \text{if } v_{i, j}^g > \bar{b}_j 
	\end{cases}.
\end{equation}

In DE, the donor vector $ \bm{v}_i^g $ and the target vector $ \bx_i^g $ mate to produce a new vector $ \bm{u}_i^g $ named the \textit{trial} vector. The binomial crossover scheme is widely used as follows:
\begin{equation} \label{eq: cx}
	u_{i, j}^g = \begin{cases}
		v_{i, j}^g  \  \text{if } \text{rand}(0, 1)  \le C_r  \text{ or }  j = \beta \\
		x_{i, j}^g \  \text{otherwise} 
	\end{cases}
\end{equation}
where $ \beta \in [1, n] $ is a random integer that is generated anew for each $ i $, and $ C_r $ is the user provided crossover rate \cite{das2011DifferentialEvolution}. {Eq.~\eqref{eq: cx} says each entry of $ \bm{u}_i^g $ comes from either $ \bm{v}_i^g $ or $ \bx_i^g $.}

Finally, DE imposes \textit{elitism} by selecting the better one between the target vector $ \bx_i^g $ and the trial vector $ \bm{u}_i^g $ as the $ i $-th vector into the next generation according to their fitness:
\begin{equation} \label{eq: selection}
	\bx_i^{g+1} = \begin{cases}
		\bm{u}_i^g  \  \text{if } f(\bm{u}_i^g) \le f(\bx_i^g) \\
		\bx_i^g  \ \text{otherwise}
	\end{cases}.
\end{equation}

\section{Experimental results and discussions}\label{sec: results}

\subsection{Datasets and experimental settings} \label{sec: datasets}
The two PV I-V datasets \cite{easwarakhanthan1986NonlinearMinimizationa} shown  in Fig. \ref{fig: BB} serve as the \textit{de facto} standard in evaluating algorithms' performance (e.g., \cite{yang2020ComprehensiveOverview, gutierrez2018SystemonChipRealTime}). The first dataset named ``RT'' contains 26 data points for an RTC France solar cell (1000 W/m$ ^2 $, 33 $ ^\circ $C). The second dataset ``PW'' with 25 data points refers to a Photowatt-PWP201 solar module (1000 W/m$ ^2 $,45 $ ^\circ $C). Combining the two datasets and two models, we consider four cases in total. The naming rule is ``\textit{model}+\textit{dataset}'' for simplicity, e.g., case ``SDM+RT'' fits the SDM to the RT dataset. The parameter search range (i.e., $\Theta$ in \eqref{eq: problem}) commonly used in the literature, as listed in Table \ref{tbl: param-range},  is adopted for fair comparisons. 

\begin{table}[tb]
	\centering
	\footnotesize
	\caption{Parameter search range in numerical experiments.}
	\label{tbl: param-range}
	\begin{tabular}{lcccc} 
		\toprule
		\multirow{2}{*}{Parameter} & \multicolumn{2}{c}{RT} & \multicolumn{2}{c}{PW}  \\ 
		\cmidrule{2-5}
		& Lower & Upper                        & Lower & Upper                               \\ 
		\midrule
		$I_{ph}$ (A)                   & 0  & 1                         & 0  & 2                                \\
		$ I_0, I_{01}, I_{02} $ ($ \mu $A)                      & 0  & 1                         & 0  & 50                               \\
		$ n, n_1, n_2 $                     & 1  & 2                         & 1  & 50                               \\
		$ R_s $  ($ \Omega $)                     & 0  & 0.5                       & 0  & 2                                \\
		$ R_p $ ($ \Omega $)                    & 0  & 100                       & 0  & 2000                             \\
		\bottomrule
	\end{tabular}
\end{table}

All metaheuristic algorithms are implemented in MATLAB R2020a for a fair comparison of runtime.  The results presented below were obtained on a laptop with a 1.8 GHz Core i7-8550U CPU, 8 GB RAM, and Windows 10. 

\subsection{Global optimality analysis via interval B\&B} \label{sec: global}
Despite the large number of metaheuristics for PV parameter estimation (see, e.g., \cite[Table 1]{calasan2020RootMean}), none of them can certify the finding of the global minimum. This section presents results regarding global optimality.

\subsubsection{SDM results}
Note that a B\&B algorithm is generally computationally intensive.
Following \cite{chenouard2020IntervalBranch}, we limited the runtime of \texttt{ibexopt} to 20000 seconds for ``SDM+RT''.  Unlike  \cite{chenouard2020IntervalBranch}, we set the absolute and relative precision to smaller values (1E-13 and 1E-9) in order to obtain tighter bounds of the global minimum (that is, $ [\underline{f}, \bar{f}] $ in Algorithm \ref{alg: BB}). The optimization results for ``SDM+RT''' are reported in the first column of Table \ref{tbl: SDM BB}. In particular, the bounds of the RMSE enjoy a negligible gap. Results reported in existing studies are mostly truncated to  five significant digits. Hence, we can \textit{certify} safely, for the first time, that 9.8602E-4 is indeed the global minimum RMSE value of ``SDM+RT''. The parameter values in Table \ref{tbl: SDM BB} yield the upper bound of RMSE, and those values match closely to results acquired with various metaheuristics (like \cite[Table 1]{calasan2020RootMean}).

\begin{table}[tb]
	\centering
	\caption{Optimization results for SDM using interval B\&B.}
	\footnotesize
	\label{tbl: SDM BB}
	\begin{tabular}{lll} 
		\toprule
		Variable & RT                                                                       & PW                                                             \\ 
		\midrule 
		$ I_{ph}  $ (A)     & 0.760779120136                                                                       & 1.03052020484                                                            \\
		$ I_0 $ ($ \mu $A)      & 0.322873926858                                                                        & 3.48287904343                                                              \\
		$ n $        & 1.48113747635                                                                        & 48.6435574734                                                            \\
		$ R_s $ ($ \Omega $)      & 0.0363792207867                                                                       & 1.20123680201                                                            \\
		$ R_p $  ($ \Omega $)       & 53.7009537057                                                                 & 981.263690780                                                              \\
		RMSE     & \begin{tabular}[c]{@{}l@{}}{[}9.860250397955652E-4,\\ \ 9.860250417458982E-4]\end{tabular} & \begin{tabular}[c]{@{}l@{}}{[}2.425076598320144E-3,\\ \ 2.425076599532477E-3]\end{tabular}  \\
		Gap      & 1.950333050615427E-12                                                                & 1.2123329007351913E-12                                                            \\
		Time (s) & 13547                                                                            & 38924                                                                \\
		\bottomrule
	\end{tabular}
\end{table}

We next examine the ``SDM+PW'' case. The major difference is the considerably widened parameter search range for PW in Table \ref{tbl: param-range}, which may pose a big challenge to the interval B\&B algorithm and require an even longer runtime. In contrast to \cite{chenouard2020IntervalBranch}, we increased the timeout to 40000 s. The best-known RMSE (see, e.g., \cite[Table 13]{long2020NewHybrid}) is \mbox{2.4250E-3}, which shows exact agreement with the RMSE bounds reported in Table \ref{tbl: SDM BB}. Besides, as expected, the parameter values reported therein are also extremely close to those in Table \ref{tbl: SDM BB}. Again, such a consensus indicates the correctness of each other. The measured and the reconstructed I-V curves using SDM and parameters in Table \ref{tbl: SDM BB} are shown in Fig.\ \ref{fig: BB} with admirable fitting accuracy. Note that the negative current and voltage values therein simply imply a reverse direction \cite{easwarakhanthan1986NonlinearMinimizationa}.

\begin{figure}[tb]
	\centering
	\begin{subfigure}{\linewidth}
		\centering
		\includegraphics[width=80mm]{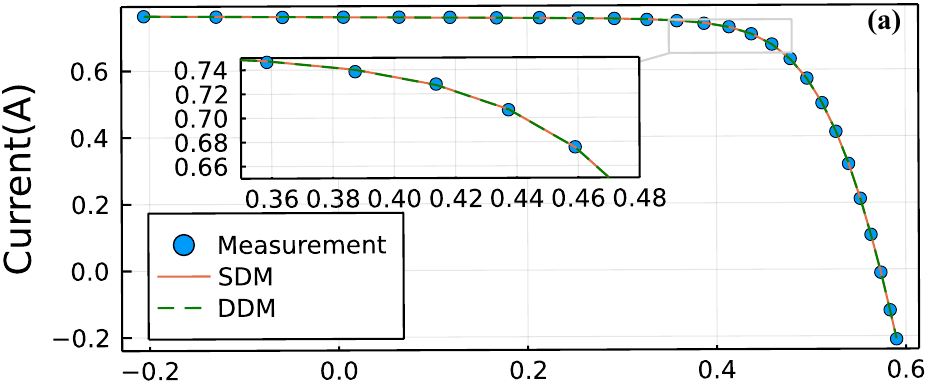} 
		\label{fig: BB_RT }
		\vspace{1mm}
	\end{subfigure}
	\\
	\begin{subfigure}{\linewidth}
		\centering
		\includegraphics[width=80mm]{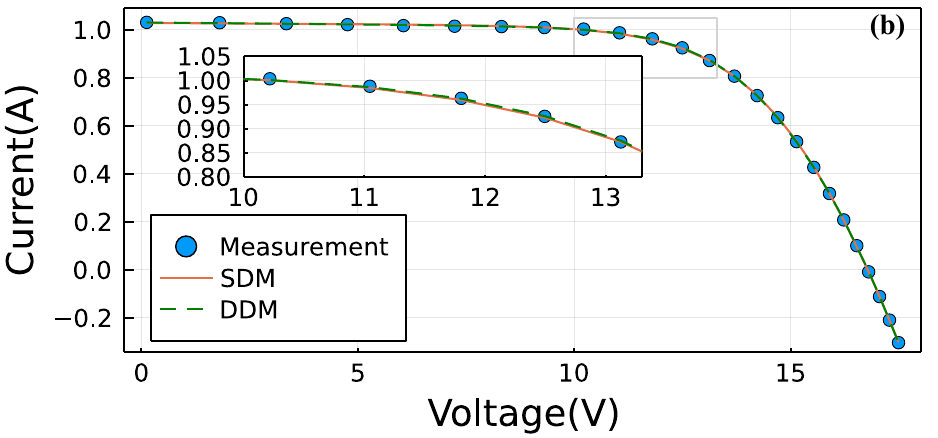} 
		\label{fig: BB_PW }
	\end{subfigure}
	\vspace*{-2mm}
	\caption{Measured and estimated I-V curves using parameter values optimized by interval B\&B. (a) RT; (b) PW. }
	\label{fig: BB}
	\vspace*{-2mm}
\end{figure} 

\subsubsection{DDM results} \label{sec: DDM BB}
The DDM is more challenging due to its two extra parameters. We decided to allow \texttt{ibexopt} more time (24 hours) to get hopefully tighter optimality bounds. The overall workflow is identical to the SDM case above. The results are reported in Table \ref{tbl: DDM BB}. Unfortunately, the lower bound remains zero, even if we run \texttt{ibexopt} for another 5 hours. This failure is probably caused by the notorious \textit{cluster effect} in B\&B methods \cite{floudas2013deterministic}. 

\begin{table}[tb]
	\centering
	\small
	\caption{Optimization results for DDM using interval B\&B.}
	\label{tbl: DDM BB}
	\begin{tabular}{lll}  
		\toprule
		Variable               & RT                                                              & PW                                              \\ 
		\midrule
		$ I_{ph} $(A)         & 0.760815738919                                                                   & 1.0339286971                                                            \\
		$ I_{01}$($\mu$A)  &0.217867184041                                                                       & 1.86575472010E-23                                                              \\
		$ I_{02}$($ \mu $A)            & 0.781454995330                                                                      & 0.535399234849                                                               \\
		$n_1$                  & 1.44827388213                                                                        & 9.58860778809                                                                     \\
		$n_2$                  &  1.98183166760                                                                       &  42.6724488388                                                                   \\
		$ R_s $($ \Omega $)   &  0.0367359827333                                                                  & 1.63619822583                                                               \\
		$ R_p $($ \Omega $)   & 55.8931982861                                                                  &  607.690281231                                                                \\
		RMSE                   & \begin{tabular}[c]{@{}l@{}}[0,\\ \ 9.83581875679E-4] \end{tabular} & \begin{tabular}[c]{@{}l@{}}[0,\\ \ 1.61865668151E-3] \end{tabular}  \\
		Gap                    &9.83581875679E-4                                                          & 1.61865668151E-3                                                            \\
		Time (s)  &   86400 & 43200    \\
		\bottomrule
	\end{tabular}
\end{table}

Despite the zero lower bound, the revealed upper bound of the RMSE is still informative since it is very close to the best-known result, e.g., 9.8248E-4 for the ``DDM+RT'' case (see \cite[Table 3]{yu2019PerformanceguidedJAYA}).  The upper bound \mbox{9.8358E-4} in Table \ref{tbl: DDM BB} implies that 9.8248E-4 is likely to be the global minimum though there is no theoretical guarantee. We did not find results reported for ``DDM+PW'' that are compatible with our analysis here. Nonetheless, the result of ``DDM+PW'' in Table \ref{tbl: DDM BB} looks reasonable by comparing with the SDM counterpart in Table \ref{tbl: SDM BB}: the RMSE upper bound of DDM is even smaller than the lower bound of SDM since, as expected, the DDM can better fit the data because of its two additional parameters \cite{liang2020ParametersEstimation}. Nonetheless, since the RMSE values of both models are pretty small, it is hard to differentiate visually the two estimated I-V curves, which are highlighted in Fig.\ \ref{fig: BB}.


\subsection{Optimization via simple differential evolution (DE)} \label{sec: opt de}	
The interval B\&B algorithm is not suitable for regular PV parameter estimation applications due to its excessively long execution time. By contrast, various metaheuristic methods can obtain a reasonably good solution in a far shorter time. However, an interesting, practically important, yet rarely studied problem is whether normal PV parameter estimation really demands the increasingly complicated metaheuristics prevailing in the recent literature. In this section, we try to get some empirical insights by examining whether the \textit{intentionally} simple DE in Algorithm \ref{alg: DE} can achieve comparable performance. Starting with the canonical values recommended in \cite[Section \RNum{3}]{das2011DifferentialEvolution}, we quickly determined appropriate control parameter values for the simple DE as $N_p = 50, C_r = 0.6, F = 0.9$, and $G=800$ (for SDM)  or $2000$  (for DDM)


Since DE is a stochastic algorithm, we follow the convention (e.g., \cite{liang2020EvolutionaryMultitask,yu2019PerformanceguidedJAYA}) to execute DE 30 times for each case and report the statistic characteristics. For illustration purposes, we list the DE results in a typical run in Table \ref{tbl: DE result}, whose estimated I-V curves are visually almost identical to Fig.\ \ref{fig: BB} (since all RMSE values are likewise tiny) and thus omitted here. The convergence curves of this simple DE for RT using both models in a typical run are shown in Fig.\ \ref{fig: CG}. The DE usually took far fewer generations to converge than the \textit{conservative} value $G$ we specified. The convergence curves of DE with the PW dataset share a similar character and is omitted here but presented in the accompanying online materials.

\begin{table}[tb]
	\centering
	\footnotesize
	\caption{Parameter values obtained by DE in a typical run.}
	\label{tbl: DE result}
	\begin{tabular}{lllll} 
		\toprule
		& \multicolumn{2}{c}{SDM}                         & \multicolumn{2}{c}{DDM}                          \\ 
		\cline{2-5}
		& \multicolumn{1}{c}{RT} & \multicolumn{1}{c}{PW} & \multicolumn{1}{c}{RT} & \multicolumn{1}{c}{PW}  \\ 
		\midrule
		$I_{ph}$(A)          & 0.760775               & 1.03051                & 0.760781               & 1.03051                 \\
		$I_0/I_{01} $$ (\mu\text{A}) $ & 0.323021               & 3.48226                & 0.225974               & 9.8113E-3              \\
		$I_{02}$($ \mu $A)            & ---                  &        ---                & 0.749344               & 3.47245                 \\
		$n/n_1$            & 1.481184               & 48.6428                & 1.45101                & 48.64282                \\
		$n_2$              &     ---                     &      ---                    &    1.99999            & 48.64283                 \\
		$R_s$($ \Omega $)              & 0.036377               & 1.20127                & 0.0367404              & 1.20127                 \\
		$R_p$($ \Omega $)                & 53.71852               & 981.982                & 55.4854                & 981.982                 \\
		RMSE               & 9.8602E-4              & 2.4250E-3              & 9.8248E-4              & 2.4250E-3               \\
		\bottomrule
	\end{tabular}
\end{table}

\begin{figure}[tb]
	\centering
	\begin{subfigure}{0.7\linewidth}
		\centering
		\includegraphics[width=60mm]{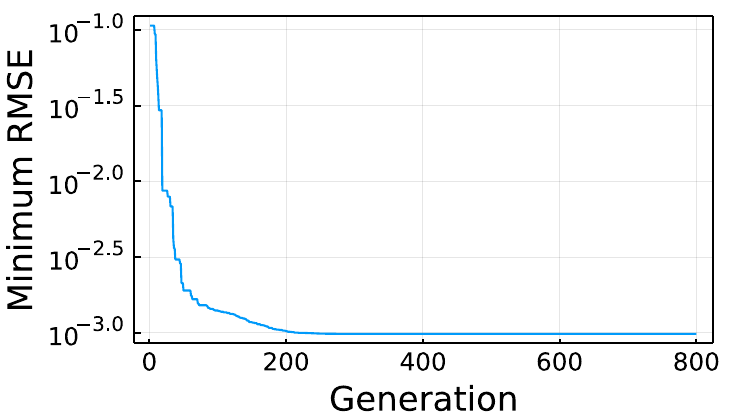} 
		\label{fig: CG_SDM }
	\end{subfigure}
	\hfill
	\begin{subfigure}{0.7\linewidth}
		\centering
		\includegraphics[width=60mm]{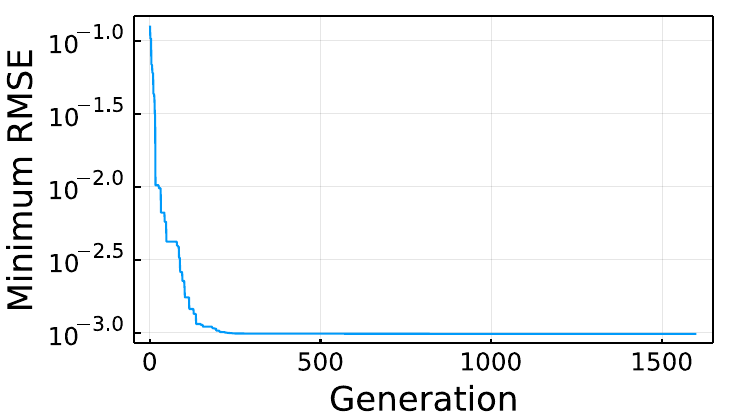} 
		\label{fig: CG_DDM }
	\end{subfigure}
	\caption{Convergence curves of DE with RT: (a) SDM (b) DDM. }
	\label{fig: CG}
\end{figure}

The statistics of RMSE in the 30 runs are reported in Table \ref{tbl: DE stats}. 
Overall, the performance of our DE algorithm is remarkably stable despite its stochasticity in nature. When applying to the SDM on both datasets and to ``DDM+PW'', the DE algorithm always yields the same minimal RMSE in all 30 trials. Even in the worst ``DDM+RT'' case, the gap between the maximum and minimum RMSE values in 30 runs is still minor, as implied by the slight standard deviation in Table \ref{tbl: DE stats}. 

\begin{table}[t]
	\centering
	\footnotesize
	\caption{Statistics of RMSE values by DE in 30 runs.}
	\label{tbl: DE stats}
	\begin{tabular}{lcccc} 
		\toprule
		& \multicolumn{2}{c}{SDM} & \multicolumn{2}{c}{DDM}  \\ 
		\cline{2-5}
		& RT         & PW         & RT        & PW           \\ 
		\midrule
		Min  & 9.8602E-4  & 2.4250E-3  & 9.8248E-4 & 2.4250E-3    \\
		Mean & 9.8602E-4  & 2.4250E-3  & 9.8267E-4 & 2.4250E-3    \\
		Max  & 9.8602E-4  & 2.4250E-3  & 9.8602E-4 & 2.4250E-3    \\
		Std  & 4.3929E-17 & 2.9525E-17 & 7.1027E-7 & 2.3955E-17   \\
		\bottomrule
	\end{tabular}
\end{table}

Table \ref{tbl: DE stats} tells  that the simple DE managed to find the global minimum RMSE, as identified by interval B\&B, for both SDM cases. As for the more challenging DDM cases, DE achieved the best-known result for ``DDM+RT'', i.e., \mbox{9.8248E-4}. However, the RMSE attained by DE for ``DDM+PW'' is still above the identified upper bound (2.4250E-3 vs 1.6186E-3 in Table~\ref{tbl: DDM BB}). The main reason is presumably attributed to the extraordinarily small value of $ I_{01} $ in the potential optimal solution (around 1.86E-23 in Table \ref{tbl: DDM BB}), which poses a overwhelming challenge to DE or any metaheuristic method in general (see Table \ref{tbl: compare} below).

\subsection{Comparison with existing algorithms}\label{sec: comp}
We compare the performance of the simple DE (Algorithm \ref{alg: DE}) with more sophisticated metaheuristics, inspecting both accuracy and efficiency. We select state-of-the-art algorithms of  distinct methodology and pick particularly DE variants for a comprehensive comparison. The existing results have been listed as they appear in four latest articles: \cite[Table 12]{li2019ParameterExtraction}, \cite[Table 3]{yu2019PerformanceguidedJAYA}, \cite[Table 9]{liang2020EvolutionaryMultitask}, and \cite[Table 9]{liang2020ParametersEstimation}.   All results are listed in Table \ref{tbl: compare}. Since the hardest case ``DDM+PW'' were not considered in these papers, we run the original source code of two recent studies \cite{yu2019PerformanceguidedJAYA, liang2020ParametersEstimation} (see \url{https://github.com/cilabzzu}) and report their results for fair comparisons. The statistics of the RMSE values in 30 runs are listed in Table~\ref{tbl: compare}. 

To examine the statistical significance of performance difference between the simple DE and the other methods, we perform the Mann-Whitney U test \cite{liang2020ParametersEstimation} and report results in the ``U test'' column of Table~\ref{tbl: compare}. The null hypothesis \(H_0\) indicates equally good performance, and the level of significance is 0.05.  In the results reported in Table~\ref{tbl: compare}, the symbol ``+" indicates a statistically significant performance difference, i.e., rejecting the null hypothesis, while ``--'' means there is no  statistically significant evidence to conclude the performance difference. 


\begin{figure}[t]
	\centering
	\includegraphics[width=80mm]{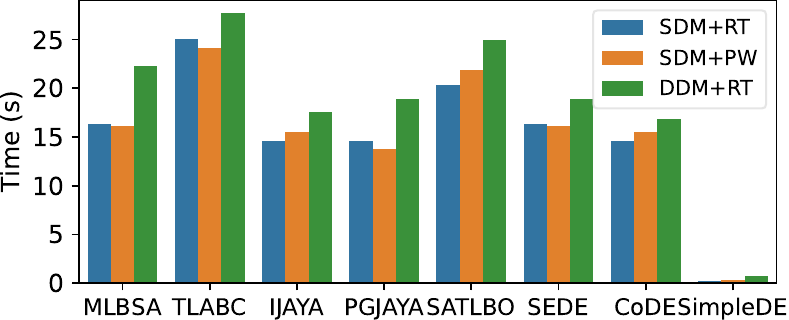} 
	\caption{Runtime comparison of different algorithms.}
	\label{fig/runtime} 
	\vspace{-2mm}
\end{figure}

\begingroup
\setlength{\tabcolsep}{2.5pt} 
\begin{table}
	\centering
	\caption{Comparison of statistical results of various algorithms in four cases. From top to bottom: ``SDM+RT'', ``SDM+PW'', ``DDM+RT'', and ``DDM+PW''. }
	\label{tbl: compare}
		\begin{tabular}{lllllc} 
			\toprule
			\multicolumn{1}{c}{\multirow{2}{*}{Method}} & \multicolumn{4}{c}{RMSE}                                                                               & \multicolumn{1}{c}{\multirow{2}{*}{{U test}}}  \\ 
			\cline{2-5}
			\multicolumn{1}{c}{}                           & \multicolumn{1}{c}{Min} & \multicolumn{1}{c}{Mean} & \multicolumn{1}{c}{Max} & \multicolumn{1}{c}{Std} & \multicolumn{1}{c}{}                           \\ 
			\midrule
			MLBSA \cite{yu2018MultipleLearning}                                          & 9.8602E-4               & 9.8602E-4                & 9.8602E-4               & 7.0800E-11              &  +                                             \\
			TLABC  \cite{chen2018TeachingLearning}                                         & 9.8602E-4               & 9.9417E-4                & 1.0308E-3               & 1.1896E-5               & +                                          \\
			IJAYA   \cite{yu2017ParametersIdentification}                                          & 9.8602E-4               & 9.8605E-4                & 9.8684E-4               & 1.4931E-7               & +                                           \\
			PGJAYA  \cite{yu2019PerformanceguidedJAYA}                                        & 9.8602E-4               & 9.8602E-4                & 9.8603E-4               & 2.8029E-9               & +                                             \\
			SATLBO \cite{yu2017ParametersIdentificationa}                                        & 9.8602E-4               & 9.8879E-4                & 1.0067E-3               & 4.8133E-6               & +                                            \\
			SGDE \cite{liang2020EvolutionaryMultitask}                                           & 9.8602E-4               & 9.8602E-4                & 9.8603E-4               & 2.4746E-9               & --                                             \\
			SEDE   \cite{liang2020ParametersEstimation}                                        & 9.8602E-4               & 9.8602E-4                & 9.8603E-4               & 4.2000E-17              & --                                            \\
			CoDE    \cite{wang2011differentialIEEE}                                          & 9.8602E-4               & 9.8602E-4                & 9.8602E-4               & \textbf{2.3100E-17}              & --                                            \\
			Simple DE                                          & \textbf{9.8602E-4 }              & \textbf{9.8602E-4 }               & \textbf{9.8602E-4}               & 2.9464E-17              &                                           \\ 
			\hline
			MLBSA                                          & 2.4250E-3               & 2.4253E-3                & 2.4336E-3               & 1.5600E-6               & +                                            \\
			TLABC                                          & 2.4250E-3               & 2.4254E-3                & 2.4287E-3               & 8.7464E-7               & +                                               \\
			IJAYA                                          & 2.4250E-3               & 2.4251E-3                & 2.4253E-3               & 5.0766E-8               & +                                               \\
			PGJAYA                                         & 2.4250E-3               & 2.4251E-3                & 2.4260E-3               & 1.7859E-7               & +                                              \\
			SATLBO                                         & 2.4250E-3               & 2.4254E-3                & 2.4315E-3               & 1.1622E-6               & +                                                \\
			SGDE                                           & 2.4250E-3               & 2.4250E-3                & 2.4250E-3               & 4.1697E-10              & +                                \\
			SEDE                                           & 2.4250E-3               & 2.4250E-3                & 2.4250E-3               & 3.1400E-17              & --                                           \\
			CoDE                                           & 2.4250E-3               & 2.4250E-3                & 2.4250E-3               & 2.1700E-17          & --                                             \\
			Simple DE                                             & \textbf{2.4250E-3}               & \textbf{2.4250E-3}                & \textbf{2.4250E-3}               & \textbf{1.7547E-17}              &                       \\ 
			\hline
			MLBSA                                          & 9.8248E-4               & 9.8506E-4                & 9.8613E-4               & 1.2400E-6               & +                                             \\
			TLABC                                          & 1.0012E-3               & 1.2116E-3                & 1.9826E-3               & 2.1100E-4               & +                                             \\
			IJAYA                                          & 9.8249E-4               & 9.8686E-4                & 9.9941E-4               & 3.2211E-6               & +                                             \\
			PGJAYA                                         & 9.8260E-4               & 9.8603E-4                & 9.9599E-4               & 2.3666E-6               & +                                             \\
			SATLBO                                         & 9.8282E-4               & 1.0054E-3                & 1.2306E-3               & 5.0271E-5               & +                                             \\
			SGDE                                           & 9.8441E-4               & 9.8577E-4                & 9.8602E-4               & \textbf{4.0150E-7}              & +                                             \\
			SEDE                                           & 9.8248E-4               & 9.8289E-4                & 9.8602E-4               & 9.1700E-7               & --                                            \\
			CoDE                                           & 9.8249E-4               & 1.0036E-3                & 1.5496E-3               & 1.0300E-4               & +                                          \\
			Simple DE                                         & \textbf{9.8248E-4}               & \textbf{9.8273E-4}                & \textbf{9.8602E-4}               & 8.9630E-7               &                                  \\ 
			\hline
			PGJAYA                                         & 2.4250E-3               & 2.4272E-3                & 2.4485E-3               & 5.4346E-6               & +                                               \\
			SEDE                                           & 2.4250E-3               & 2.4250E-3                & 2.4250E-3               & 6.6661E-17              & --                                               \\
			Simple DE                                            & \textbf{2.4250E-3}               & \textbf{2.4250E-3}                & \textbf{2.4250E-3 }              & \textbf{2.7356E-17}              &                                      \\
			\bottomrule
		\end{tabular}
		\vspace{-2mm}
\end{table}
\endgroup

Overall, our simple DE (Algorithm \ref{alg: DE}) and several strong competitors like SEDE can attain the best RMSE values.  Statistically, the Mann-Whitney U test indicates that the accuracy of the  simple DE is on par with the selected state-of-the-art approaches such as SEDE. The result is somewhat surprising given the extreme simplicity of the simple DE. Unfortunately, such simple metaheuristics have been largely overlooked in the current literature. 
Note that we compare the simple DE intentionally with three more complicated DE variants: SGDE, SEDE, and CoDE. In the two SDM cases in Table~\ref{tbl: compare}, the four DE algorithms exhibit almost identical accuracy in terms of RMSE values, while CoDE demonstrates the highest stability measured by the standard deviation and SGDE is the least stable one. By contrast, in the more challenging DDM cases, our simple DE outperforms both CoDE and SEDE with its enhanced performance stability. 

The most apparent advantage of the simple DE is its substantially reduced running time ($< 1$ s). This impressive speedup is mainly brought by its extreme simplicity including only four computationally cheap equations in Algorithm \ref{alg: DE}. {Besides, note that the evaluation of the objective function \eqref{eq: problem obj} with a few dozens of data points is inexpensive, which implies consequently that it is usually the algorithm's internal computation burden that dominates the overall time consumption.} This claim is supported particularly by the significantly longer runtime of the other three more complex DE variants in Fig.~\ref{fig/runtime}. 

As for the ``DDM+PW'' case in Table \ref{tbl: compare},  we notice that the best RMSE value attained by the three algorithms all turns out to be 2.4250E-3, though this value is certainly not the global minimum (recall the upper bound ascertained in Table~\ref{tbl: DDM BB}).  As mentioned in Section \ref{sec: opt de}, this failure is possibly caused by the excessively small true value of $ I_{01} $ (see Table \ref{tbl: DDM BB}) that can challenge all metaheuristic methods.

In Table~\ref{tbl: compare}, the RMSE values attained by various methods, on the level of 1E-4 or 1E-3, seem sufficiently small for practical applications. Such observations also justify the use of a simpler algorithm from another angle: the excessively high accuracy may be pragmatically unnecessary, and a practitioner can opt to trade off accuracy with algorithmic simplicity. Even better, extensive examinations above have validated the competitive accuracy of the simple DE method despite its extraordinary simplicity and efficiency. 


\section{Conclusion}\label{sec: conclusion}
In this paper, we tried to address two essential issues of PV parameter estimation that have seldom been attempted in the current literature. With the two most widely used benchmark datasets, the globally minimum RMSE for the SDM and a reasonably tight upper bound for the DDM were certified rigorously by an interval analysis based B\&B algorithm. However, the running time of this interval B\&B algorithm is overly long for practical usages despite its theoretical guarantee. Next, we showed through extensive examination that, for the first time and somewhat surprisingly, a simple DE algorithm (Algorithm \ref{alg: DE}) was capable of locating the global minimum or at least attaining the best-known result. Moreover, the simple {and easy-to-tune} DE algorithm is distinguished by its favorable performance stability and unmatched efficiency. 
Our findings imply that, unfortunately, many existing metaheuristics for PV parameter estimation might be overly complicated and risk over-engineering. 
We suggest that a practitioner start with the simple DE as the off-the-shelf tool, especially in real-time parameter estimation scenarios.  Our code is available at \url{https://github.com/ShuhuaGao/rePVest} to ease reproduction. 



\bibliographystyle{IEEEtran}
\bibliography{./Bibliography/PVestimation}\ 

\begin{thebibliography}{10}
\providecommand{\url}[1]{#1}
\csname url@samestyle\endcsname
\providecommand{\newblock}{\relax}
\providecommand{\bibinfo}[2]{#2}
\providecommand{\BIBentrySTDinterwordspacing}{\spaceskip=0pt\relax}
\providecommand{\BIBentryALTinterwordstretchfactor}{4}
\providecommand{\BIBentryALTinterwordspacing}{\spaceskip=\fontdimen2\font plus
\BIBentryALTinterwordstretchfactor\fontdimen3\font minus
  \fontdimen4\font\relax}
\providecommand{\BIBforeignlanguage}[2]{{%
\expandafter\ifx\csname l@#1\endcsname\relax
\typeout{** WARNING: IEEEtran.bst: No hyphenation pattern has been}%
\typeout{** loaded for the language `#1'. Using the pattern for}%
\typeout{** the default language instead.}%
\else
\language=\csname l@#1\endcsname
\fi
#2}}
\providecommand{\BIBdecl}{\relax}
\BIBdecl

\bibitem{jordehi2016ParameterEstimation}
A.~R. Jordehi, ``Parameter estimation of solar photovoltaic ({{PV}}) cells:
  {{A}} review,'' \emph{Renewable and Sustainable Energy Reviews}, vol.~61, pp.
  354--371, Aug. 2016.

\bibitem{li2019ParameterExtraction}
S.~Li, W.~Gong, X.~Yan, C.~Hu, D.~Bai, L.~Wang, and L.~Gao, ``Parameter
  extraction of photovoltaic models using an improved teaching-learning-based
  optimization,'' \emph{Energy Conversion and Management}, vol. 186, pp.
  293--305, Apr. 2019.

\bibitem{yang2020ComprehensiveOverview}
B.~Yang, J.~Wang, X.~Zhang, T.~Yu, W.~Yao, H.~Shu, F.~Zeng, and L.~Sun,
  ``Comprehensive overview of meta-heuristic algorithm applications on {{PV}}
  cell parameter identification,'' \emph{Energy Conversion and Management},
  vol. 208, p. 112595, Mar. 2020.

\bibitem{villalva2009ComprehensiveApproach}
M.~G. Villalva, J.~R. Gazoli, and E.~R. Filho, ``Comprehensive {{Approach}} to
  {{Modeling}} and {{Simulation}} of {{Photovoltaic Arrays}},'' \emph{IEEE
  Transactions on Power Electronics}, vol.~24, no.~5, pp. 1198--1208, May 2009.

\bibitem{chenouard2020IntervalBranch}
R.~Chenouard and R.~A. {El-Sehiemy}, ``An interval branch and bound global
  optimization algorithm for parameter estimation of three photovoltaic
  models,'' \emph{Energy Conversion and Management}, vol. 205, p. 112400, Feb.
  2020.

\bibitem{chin2015CellModelling}
V.~J. Chin, Z.~Salam, and K.~Ishaque, ``Cell modelling and model parameters
  estimation techniques for photovoltaic simulator application: {{A}} review,''
  \emph{Applied Energy}, vol. 154, pp. 500--519, Sep. 2015.

\bibitem{nunes2018NewHigha}
H.~G.~G. Nunes, J.~A.~N. Pombo, S.~J. P.~S. Mariano, M.~R.~A. Calado, and
  J.~A.~M. {Felippe de Souza}, ``A new high performance method for determining
  the parameters of {{PV}} cells and modules based on guaranteed convergence
  particle swarm optimization,'' \emph{Applied Energy}, vol. 211, pp. 774--791,
  Feb. 2018.

\bibitem{yu2017ParametersIdentification}
K.~Yu, J.~J. Liang, B.~Y. Qu, X.~Chen, and H.~Wang, ``Parameters identification
  of photovoltaic models using an improved {{JAYA}} optimization algorithm,''
  \emph{Energy Conversion and Management}, vol. 150, pp. 742--753, Oct. 2017.

\bibitem{yu2019PerformanceguidedJAYA}
K.~Yu, B.~Qu, C.~Yue, S.~Ge, X.~Chen, and J.~Liang, ``A performance-guided
  {{JAYA}} algorithm for parameters identification of photovoltaic cell and
  module,'' \emph{Applied Energy}, vol. 237, pp. 241--257, Mar. 2019.

\bibitem{liang2020ParametersEstimation}
J.~Liang, K.~Qiao, K.~Yu, S.~Ge, B.~Qu, R.~Xu, and K.~Li, ``Parameters
  estimation of solar photovoltaic models via a self-adaptive ensemble-based
  differential evolution,'' \emph{Solar Energy}, vol. 207, pp. 336--346, Sep.
  2020.

\bibitem{chen2018TeachingLearning}
X.~Chen, B.~Xu, C.~Mei, Y.~Ding, and K.~Li, ``Teaching\textendash
  learning\textendash based artificial bee colony for solar photovoltaic
  parameter estimation,'' \emph{Applied Energy}, vol. 212, pp. 1578--1588, Feb.
  2018.

\bibitem{long2020NewHybrid}
W.~Long, S.~Cai, J.~Jiao, M.~Xu, and T.~Wu, ``A new hybrid algorithm based on
  grey wolf optimizer and cuckoo search for parameter extraction of solar
  photovoltaic models,'' \emph{Energy Conversion and Management}, vol. 203, p.
  112243, 2020.

\bibitem{gutierrez2018SystemonChipRealTime}
R.~Guti{\'e}rrez, J.~M. Blanes, D.~Marroqu{\'i}, A.~Garrig{\'o}s, and F.~J.
  Toledo, ``System-on-{{Chip}} for {{Real}}-{{Time Satellite Photovoltaic
  Curves Telemetry}},'' \emph{IEEE Transactions on Industrial Informatics},
  vol.~14, no.~3, pp. 951--957, Mar. 2018.

\bibitem{easwarakhanthan1986NonlinearMinimizationa}
T.~Easwarakhanthan, J.~Bottin, I.~Bouhouch, and C.~Boutrit, ``Nonlinear
  minimization algorithm for determining the solar cell parameters with
  microcomputers,'' \emph{International journal of solar energy}, vol.~4,
  no.~1, pp. 1--12, 1986.

\bibitem{liang2020EvolutionaryMultitask}
J.~Liang, K.~Qiao, M.~Yuan, K.~Yu, B.~Qu, S.~Ge, Y.~Li, and G.~Chen,
  ``Evolutionary multi-task optimization for parameters extraction of
  photovoltaic models,'' \emph{Energy Conversion and Management}, vol. 207, p.
  112509, Mar. 2020.

\bibitem{floudas2013deterministic}
C.~A. Floudas, \emph{Deterministic global optimization: theory, methods and
  applications}.\hskip 1em plus 0.5em minus 0.4em\relax Springer Science \&
  Business Media, 2013, vol.~37.

\bibitem{hansen2003global}
E.~Hansen and G.~W. Walster, \emph{Global optimization using interval analysis:
  revised and expanded}.\hskip 1em plus 0.5em minus 0.4em\relax CRC Press,
  2003, vol. 264.

\bibitem{das2011DifferentialEvolution}
S.~Das and P.~N. Suganthan, ``Differential {{Evolution}}: {{A Survey}} of the
  {{State}}-of-the-{{Art}},'' \emph{IEEE Transactions on Evolutionary
  Computation}, vol.~15, no.~1, pp. 4--31, Feb. 2011.

\bibitem{calasan2020RootMean}
M.~{\'C}alasan, S.~H.~E. Abdel~Aleem, and A.~F. Zobaa, ``On the root mean
  square error ({{RMSE}}) calculation for parameter estimation of photovoltaic
  models: {{A}} novel exact analytical solution based on {{Lambert W}}
  function,'' \emph{Energy Conversion and Management}, vol. 210, p. 112716,
  Apr. 2020.

\bibitem{yu2018MultipleLearning}
K.~Yu, J.~J. Liang, B.~Y. Qu, Z.~Cheng, and H.~Wang, ``Multiple learning
  backtracking search algorithm for estimating parameters of photovoltaic
  models,'' \emph{Applied Energy}, vol. 226, pp. 408--422, Sep. 2018.

\bibitem{yu2017ParametersIdentificationa}
K.~Yu, X.~Chen, X.~Wang, and Z.~Wang, ``Parameters identification of
  photovoltaic models using self-adaptive teaching-learning-based
  optimization,'' \emph{Energy Conversion and Management}, vol. 145, pp.
  233--246, Aug. 2017.

\bibitem{wang2011differentialIEEE}
Y.~Wang, Z.~Cai, and Q.~Zhang, ``Differential evolution with composite trial
  vector generation strategies and control parameters,'' \emph{IEEE
  transactions on evolutionary computation}, vol.~15, no.~1, pp. 55--66, 2011.

\end{thebibliography}

\end{document}